\documentstyle[aps,multicol,epsf]{revtex}
\input epsf
\draft
\title{High precision atom interferometry in a microgravity environment}
\author{Tycho Sleator
\\ {\em New York University, Physics Department, 4 Washington Pl., New York, NY 10003} }
\author{Paul R. Berman and Boris Dubetsky
\\ {\em University of Michigan, Physics Department, Ann Arbor, MI 48109-1120} }
\date{\today }
\begin{document}
\maketitle
\begin{abstract}
We propose a set of experiments in which Ramsey-fringe techniques are
tailored to probe transitions originating and terminating on the same ground
state level. When pulses of resonant radiation, separated by a time delay $%
T, $ interact with atoms, it is possible to produce Ramsey fringes having
widths of order $1/T$. If each pulse contains two counterpropagating
travelling wave modes, the atomic wave function is split into two or more
components having different center-of-mass momenta. Matter-wave interference
of these components leads to atomic gratings, which have been observed in
both spatially separated fields and time separated fields. Time-dependent
signals can be transformed into frequency dependent signals, leading to
ground state Ramsey fringes (GSRF). The signals can be used to probe many
problems of fundamental importance:

\ - If a gravitational field is present, the atomic gratings acquire an
additional phase shift. For ground based experiments the shift can be used
to obtain a precise measurement of the gravitational acceleration $g$. For
experiments in a microgravity environment, GSRF can be used to measure
residual gravitational fields that can be as small as $6\,10^{-9}g.$

\ - When atoms move in a rotating frame, the atomic grating acquires an
additional phase caused by Coriolis acceleration, which also leads to a
shift of the GSRF line-center. We propose to use GSRF as the basis for a new
type of gyroscope. Very preliminary estimates show that in a microgravity
environment one can measure a rotation rate with an accuracy of $6\,10^{-3}%
{{}^\circ}%
/h$, which is 10 times better than that achieved using a fiber optic
gyroscope.

\ - Since atomic scattering from the pulses is accompanied by a momentum
change, i.\ e.\ by recoil, a modulation of the grating dependence with time
delay $T$ occurs. This modulation, whose frequency is equal to the atomic
recoil frequency, leads to recoil splitting of the GSRF signal. The recoil
splitting can be resolved with relative accuracy $10^{-6}$ and used for
recoil frequency measurements, important for a precise determination of
Planck's constant.

\ Since only transitions originating and terminating on the same ground
state are involved, frequency measurements can be carried out using lasers
phase-locked by quartz oscillators having relatively low frequency. Our
technique may allow one to increase the precision by a factor of 100 (the
rf- to quartz oscillator frequencies ratio) over previous experiments based
on Raman-Ramsey fringes or reduce on the same factor requirements for
frequency stabilization.
\end{abstract}
\begin{multicols}{2}
\narrowtext

\section{Introduction}

Recent progress in cooling and trapping of neutral atoms allows one to
observe extremely slow processes involving atoms in their ground states.
These processes can serve as the basis for a new generation of atomic clocks
whose operation is based entirely or partially on matter-wave interferometry.

For atoms having a velocity spread of order $1\,$cm/s that are confined to
an interaction volume of $1\,$cm$^{3},$ one can observe the evolution of
various ground-state coherences for times $T_{e}$ of order $1\,$s$.$ The
line widths associated with these coherences can be as small as $0.15\,$Hz,
allowing for frequency measurements having an accuracy of order $0.01-1\,$mHz%
$.$ Very often, the earth's gravitational field is the limiting factor that
determines the accuracy one can achieve in these measurements. The most
important effect of the Earth's gravitational field is to introduce a
Doppler frequency shift in matter-radiation field interactions. These shifts
can be significant in high precision measurements of atomic recoil or
rotational sensing. For experiments carried out on the time scale of $1\,s$,
atoms are accelerated in the Earth's field to a velocity of $10^{3}$ cm/s,
which leads to a Doppler shift on the order of $10$ MHz. This large Doppler
shift can often obscure the small recoil or rotational shifts one is
attempting to measure in high precision experiments. {\em As a result, it
would be useful and important to use a microgravity environment to carry out
high precision measurements of fundamental constants and inertial effects,
where such measurements could be carried out with unprecedented precision}.
In addition to the advantages gained by elimination of the gravitational
Doppler shift, one also finds that the atom-field interaction time can be
increased since atoms do not fall out of the atom-radiation field
interaction region as a result of gravitational acceleration.

In this article we discuss a new technique for precision measurements based
entirely on matter wave interference\cite{0}, where one produces coherences
between different atomic center-of-mass states within the same internal
state. For the current state of atom interferometry, see \cite{0p}. The
exclusion of any internal transitions allows one either to increase the
measurement accuracy, or, for a fixed accuracy, to reduce the requirements
for frequency stabilization.

The article is arranged as follows. In the next section we compare briefly
different techniques to estimate the advantages of the matter wave
interference method. In sections III and IV we present our previous results
for gravitational acceleration and recoil frequency measurements. The ground
state Ramsey fringe technique (GSRF) is discussed in Sec. V. In Sec. VI we
estimate the residual gravity measurement in the microgravity environment.
Sec. VII is devoted to an atom gyroscope in a microgravity environment and
in a ground-based experiment. One possibility for increasing experimental
precision by producing higher order atom gratings is discussed in Sec. VIII.
An observation of these gratings using a 3-pulse echo technique is presented
in Sec. IX. The results are summarized in Sec. X.

\section{Comparison of the different techniques}

Even in a microgravity environment the time of evolution $T_{e}$ is still
restricted by the instability of the laser frequency that produces this
coherence. If the coherence oscillates at a frequency $\omega _{e}$ with
stability $\alpha ,$ then evidently the limitation on $T_{e}$ is given by 
\begin{equation}
T_{e}\lesssim \left( \alpha \omega _{e}\right) ^{-1}.  \label{t1}
\end{equation}
Typically, the Ramsey technique \cite{1} is used for precise measurements,
where one applies two or more resonant pulses separated by a time delay $%
T\sim T_{e}.$ If a resonant optical field having frequency $\Omega $ and
wave vector ${\bf k}$ (see fig. \ref{f01}a) excites an atom from the initial
state $\left| i,{\bf p}\right\rangle $ to the final state $\left| f,{\bf p+}%
\hbar {\bf k}\right\rangle ,$ where ${\bf p}$ and ${\bf p+}\hbar {\bf k}$
are the initial and final atomic center-of-mass momenta, and $i$ and $f$
label internal degrees of freedom, the atom coherence evolves in free space
at a frequency 
\begin{equation}
\omega _{atom}=\frac{\varepsilon _{f,{\bf p+}\hbar {\bf k}}-\varepsilon _{i,%
{\bf p}}}{\hbar }=\omega _{fi}+{\bf k}\cdot {\bf v+}\omega _{k},  \label{t2}
\end{equation}
where the state energy $\varepsilon _{f,{\bf p}}$ consists of the internal
energy $\varepsilon _{f}$ and kinetic energy $p^{2}/2m,$ $m$ is an atom
mass, $\omega _{fi}$ is a transition frequency, ${\bf v}={\bf p/}m$ is an
atomic velocity, $\omega _{k}=\hbar k^{2}/2m$ is a recoil frequency. For a
time $T$ between successive pulses, the atom coherence and field acquire
different phases, $\omega _{atom}T$ and $\Omega T$, respectively. For the
single-photon Ramsey fringe technique sensitive to this phase difference,
this leads to an atom- field dephasing $\phi $ that consists of three parts, 
\cite{1p} 
\begin{mathletters}
\label{t3}
\begin{eqnarray}
\phi  &=&\phi _{R}-\phi _{D}-\phi _{q},  \label{t3a} \\
\phi _{R} &=&\Delta T,  \label{t3b} \\
\phi _{D} &=&{\bf k}\cdot {\bf v}T,  \label{t3c} \\
\phi _{q} &=&\omega _{k}T,  \label{t3d}
\end{eqnarray}
where $\Delta =\Omega -\omega _{fi}$ is the atom-field detuning. These three
parts play qualitatively different roles. 
\begin{figure}[tb!]
\centering
\begin{minipage}{8.0cm}
\epsfxsize= 8 cm \epsfysize= 3.67 cm \epsfbox{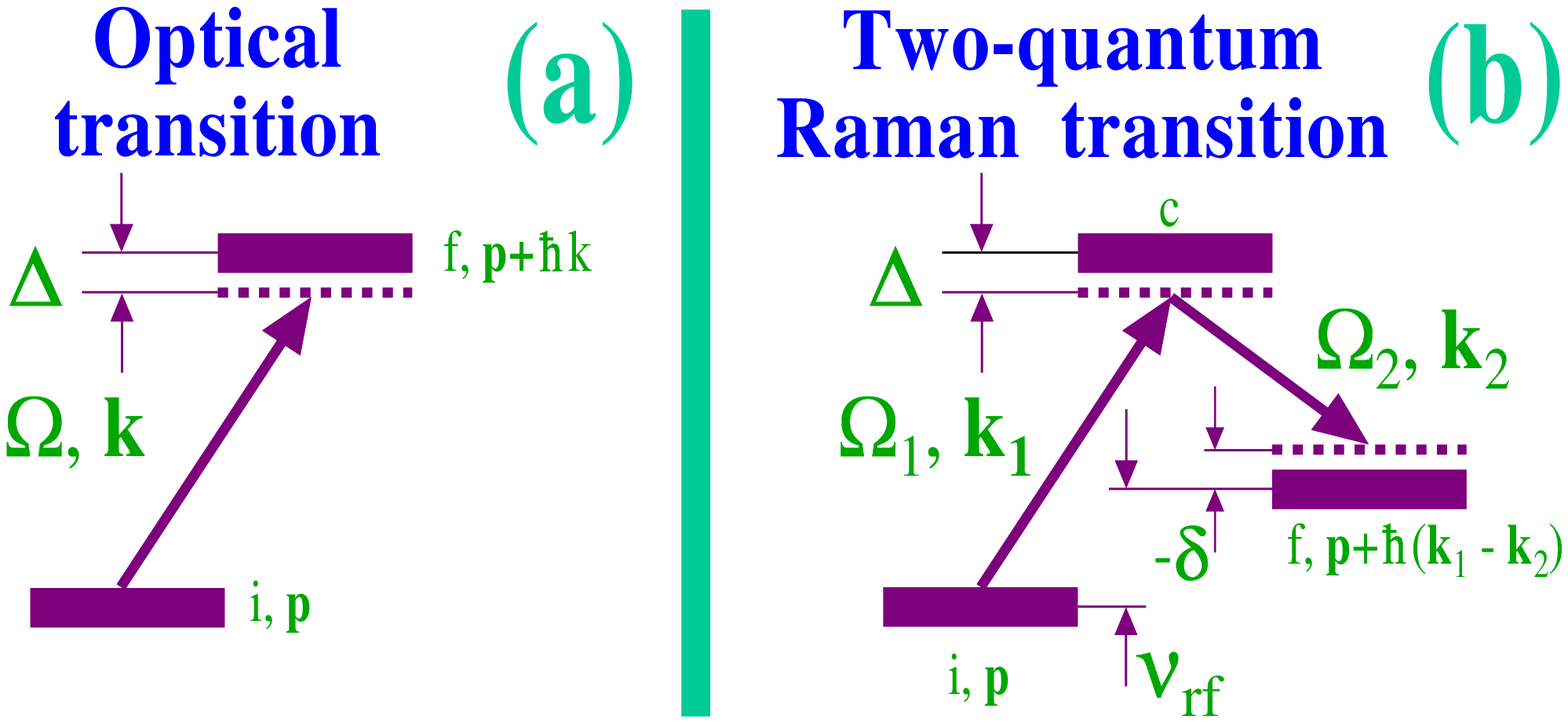}
\end{minipage}
\caption{Comparison between one photon (a) and Raman (b) techniques.}
\label{f01}
\end{figure}
The Ramsey phase $\phi _{R}$ leads to Ramsey fringes which one uses for
frequency stabilization to create precise atomic clocks. For a given measure
of the frequency stability $\alpha ,$ the limit for the time evolution of
the coherence is given by 
\end{mathletters}
\begin{equation}
T_{e}\lesssim T_{opt}\sim \left( \alpha \Omega \right) ^{-1}  \label{t4}
\end{equation}

In the {\em rf} domain the Doppler phase, $\phi _{D}\ll 1$, can be
neglected. On the other hand, in the optical domain, $\phi _{D}>1$ and, on
averaging over the velocity distribution, the Ramsey fringe signal is
destroyed\cite{2}. The washing out of the signal is analogous to that which
occurs in free induction decay (FID). The signal can be restored by the use
of echo techniques involving multiple pulses of standing-wave fields\cite{3}
and/or counterpropagating optical fields. (If one uses copropagating
traveling waves, the Ramsey phase and the Doppler phase both go to zero at
the time the echo signal is generated.) Optical Ramsey fringes using three
spatially separated standing waves were observed first in Ne on the
transition at $\lambda =0.5882$ $\mu $ \cite{4}.

In contrast to the washing out of the macroscopic atom coherence, the
Doppler phase can play a positive role in the optical domain. When the atom
velocity changes owing to external forces, a total cancellation of the
Doppler phase does not occur. If the atom acceleration is independent of the
projection of the velocity along the wave propagation direction, the
residual Doppler dephasing does not wash out the Ramsey fringes, but it can
lead to a shift or deformation of the Ramsey fringe line shape. Therefore,
optical Ramsey fringes can serve as a sensor of atomic acceleration\cite{5}.
They were used first for this purpose in Ca on the transition at $\lambda
=0.6527$ $\mu m$ \cite{6} to measure a rotation frequency.

Owing to multiphoton processes in an atom's interaction with a standing wave
field, quantum dephasing (\ref{t3d}) is responsible for the recoil splitting
of optical Ramsey fringes\cite{7} into components centered at 
\begin{equation}
\Delta \sim \pm \omega _{k},  \label{t5}
\end{equation}
first observed also on the transition in Ca \cite{8}.

It has proven advantageous to substitute a two-photon Raman transition
between different atomic ground state hyperfine sublevels \cite{30} for the
one-photon optical transitions (see fig. \ref{f01}b). Consider the
interaction of an atom with two optical waves having frequencies $\Omega
_{1} $ and $\Omega _{2}$, both nearly resonant with the coupled atomic
transitions $i\rightarrow c$ and $c\rightarrow f,$ respectively, where $i$
and $f$ are the initial and final hyperfine sublevels of the atomic ground
state. When the detunings, $\Delta _{1}$ and $\Delta _{2}$, are larger than
the inverse pulse duration, it is possible to drive a two-quantum transition
between states $i$ and $f$ which is resonant when

\begin{equation}
\delta =\Omega _{2}-\Omega _{1}+\nu _{rf}  \label{19p}
\end{equation}
is equal to 0. When an atom interacts with separated pulses, where each
pulse consists of a pair of fields having frequencies $\Omega _{1}$ and $%
\Omega _{2}$, the Ramsey phase becomes 
\begin{equation}
\phi _{R}=\delta T.  \label{19pp}
\end{equation}

If ${\bf k}_{1}$ and ${\bf k}_{2}$ are wave vectors of the fields,
absorption from mode $\left( \Omega _{1},\,{\bf k}_{2}\right) $ and
stimulated emission into mode $\left( \Omega _{2},\,{\bf k}_{2}\right) ,$
leads to a momentum change for the atoms given by ${\bf p}\rightarrow {\bf p}%
+\hbar {\bf q,}$ where 
\begin{equation}
{\bf q=k}_{1}{\bf -k}_{2}.  \label{19ppp}
\end{equation}

The atomic coherence $i\rightarrow f$ oscillates now at the frequency 
\begin{equation}
\omega _{atom}=\frac{\varepsilon _{f,{\bf p+}\hbar {\bf q}}-\varepsilon _{i,%
{\bf p}}}{\hbar }.  \label{19}
\end{equation}
From a comparison with Eq. (\ref{t2}), one concludes that for a qualitative
consideration of Ramsey fringes on two-quantum transitions, one can simply
substitute 
\begin{equation}
{\bf k\rightarrow q=k_{1}-k_{2}}.  \label{20}
\end{equation}
in Eqs. (\ref{t3c}, \ref{t3d}) to obtain Doppler and quantum dephasings, 
\begin{mathletters}
\label{t5p}
\begin{eqnarray}
\phi _{D} &=&{\bf q}\cdot {\bf v}T,  \label{t5pa} \\
\phi _{q} &=&\omega _{q}T,  \label{t5pb}
\end{eqnarray}
respectively.

The effective wave vector $q$ changes from $0$ for copropagating waves to $%
2k $ for counterpropagating waves. Consequently, dephasing can be equal to
or greater than the dephasing for optical Ramsey fringes. Shifts of the line
center caused by quantum, inertial, or gravitational effects are also of the
same order of magnitude. However, the shifts are now measured relative to
the rf transition frequency, $\nu _{rf}$. Even though the line width of each
laser oscillating at the frequencies $\Omega _{1}$ and $\Omega _{2}$ could
be of order of 1 MHz, modern stabilization techniques allow \cite{31} one to
control the frequency difference, $\Omega _{1}-\Omega _{2}$, with a
precision $\alpha \nu _{rf},$ and, therefore, the evolution time of the
coherence is restricted by 
\end{mathletters}
\begin{equation}
T_{e}\lesssim T_{Raman}\sim \left( \alpha \nu _{rf}\right) ^{-1}  \label{t6}
\end{equation}
Compared with the evolution time (\ref{t4}) allowed by the optical
transition, the time (\ref{t6}) can be up to a factor 
\begin{equation}
A\sim \frac{T_{Raman}}{T_{opt}}\sim \frac{\Omega }{\nu _{rf}}\sim 10^{4}\ to%
\text{ }10^{5}.  \label{23}
\end{equation}
larger, which allows one to increase tremendously the accuracy of precision
measurements.

The Raman-Ramsey technique was used \cite{32} to measure the recoil
frequency $\omega _{q}$ in Cs (transition $\nu _{rf}\approx 9.2\,$GHz) with
an accuracy $1.1\times 10^{-7}$. This measurement is important for a precise
determination of the fine structure constant \cite{32}. A sensitivity to
gravitational acceleration at the level $3\times 10^{-8}g$ has been observed 
\cite{20} using the Raman-Ramsey resonance in sodium atoms (transition $\nu
_{rf}\approx 1.7\,$GHz). An atom gyroscope operating on the transition in Cs
has been created\cite{Gus97} as well, which can have a long term stability
of $10^{-5}$ $%
{{}^\circ}%
/h$ for rotation measurements \cite{t9}.

\begin{figure}[tb!]
\centering
\begin{minipage}{8.0cm}
\epsfxsize= 8 cm \epsfysize= 5.6 cm \epsfbox{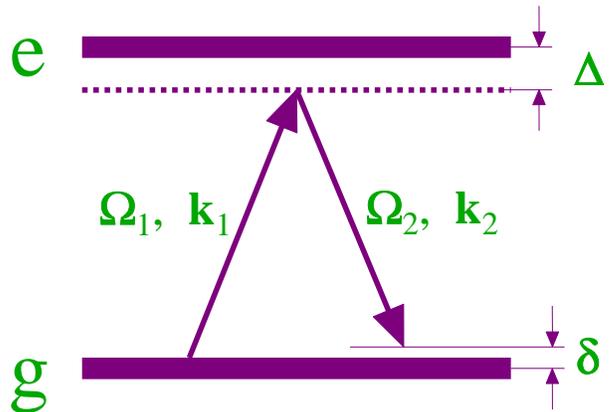}
\end{minipage}
\caption{Matter wave interference scheme.}
\label{f02}
\end{figure}

In the progression from ''one-photon optical Ramsey fringes'' to
''two-quantum Raman-Ramsey fringes'', we propose the next natural step for
experiments with fields separated in time or space. Only optical transitions 
$i\rightarrow c$ and $c\rightarrow f$ are responsible for the Doppler and
recoil effects, while the transition $i\rightarrow f$ is not relevant to
these effects. On the other hand the $i\rightarrow f$ transition frequency
still restricts the time of evolution. It cannot be decreased out of the
microwave range since atom hyperfine transition frequencies belong to this
region. The restriction can only be relaxed \cite{t10} if both the final and
initial internal states coincide (see fig. \ref{f02}). In this case only a
transition between different center-of-mass states, ${\bf p}\ $and ${\bf p}%
+n\hbar {\bf q}$, occurs. According to a classification of atom
interferometers,\cite{t11} this scheme belongs the class, matter wave atom
interferometers (MWAI).
There is no inherent reason why MWAI would produce worse measurement errors
than the Raman-Ramsey technique, but owing to the absence of any internal
transitions, the MWAI has the potential to provide more precise
measurements. Indeed, the interference between initial and final states 
\begin{equation}
\Psi _{in}\propto \exp \left( i{\bf p}\cdot {\bf r/}\hbar \right) \text{ and 
}\Psi _{out}\propto \exp \left[ i\left( {\bf p+}\hbar {\bf q}\right) \cdot 
{\bf r/}\hbar \right]  \label{t7}
\end{equation}
leads to a grating in the atomic ground state population 
\begin{equation}
\rho _{gg}\left( {\bf r}\right) \propto \cos \left( {\bf q}\cdot {\bf r}%
\right) .  \label{t8}
\end{equation}
We propose to use this grating as a sensor of the gravity, inertial, and
quantum effects, instead of the coherence $i\rightarrow f$ in the
Raman-Ramsey scheme. The quantum dephasing (\ref{t5pb}) affects only the
population grating amplitude \cite{0}. The velocity-dependent part of the
Doppler phase (\ref{t5pa}) is canceled at an echo time, and by avoiding
internal transitions, we exclude any restrictions on increasing the grating
evolution time $T_{e}$ that might arise from the physics of the atom-field
interaction.

\begin{figure}[tb!]
\centering
\begin{minipage}{8.0cm}
\epsfxsize= 8 cm \epsfysize= 1.98 cm \epsfbox{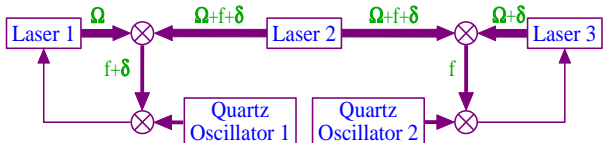}
\end{minipage}
\caption{To detune lasers 1 and 3 by a small detuning $\protect\delta $ one
can use an intermediate laser 2, detune it from laser 1 by the
frequency $f+\protect\delta $ and control this detuning by a quartz
oscillator 1. After that one detunes laser 3 from laser 2 by a frequency 
-$f$ controlled by the quartz oscillator 2.}
\label{f03}
\end{figure}

The restriction on $T_{e}$ is rather technical. A small detuning $\delta $
(of order of several mHz) definitely resides within the noise bandwidth of
the fields $\left( \Omega _{1},{\bf k}_{1}\right) $ and $\left( \Omega _{2},%
{\bf k}_{2}\right) $. Therefore, these fields cannot be detuned by $\delta $
directly. To detune them, one can use highly stabilized quartz oscillators 1
and 2 operating on the frequencies $f$ and $f+\delta $, respectively, where $%
f$ is a typical quartz oscillator frequency. The only requirement in
choosing $f$ is that it be larger than the lasers' noise bandwidth. A
sequence of detunings of lasers 1, 2 and 3, as shown in fig. \ref{f03},
controlled by quartz oscillators 1 and 2, solves the problem.
For this scheme\ the evolution time of the grating is restricted by the
quartz oscillator frequency instability $\alpha f,$%
\begin{equation}
T_{e}\lesssim T_{MWAI}\sim \left( \alpha f\right) ^{-1}.  \label{t9}
\end{equation}
Comparing this result with the inequality (\ref{t6}), one finds that using
the same initial and final states allows one to increase the measurement
time by a factor 
\begin{equation}
A\sim \frac{T_{MWAI}}{T_{Raman}}\sim \frac{\nu _{rf}}{f}\sim 10^{2}-10^{3},
\label{t10}
\end{equation}
where we assume for $f$ a typical value $\left( 10-100\,MHz\right) .$ Even
if for other reasons, such as a wave front curvature or magnetic field
gradient, a further increase of the evolution time becomes impossible, our
scheme still has an advantage. For a given $T_{e}$ using MWAI, one has to
have the frequency stability 
\begin{equation}
\alpha \lesssim fT_{e},  \label{t11}
\end{equation}
while for the Raman-Ramsey technique one has to stabilize frequency to $%
\alpha \lesssim \nu _{rf}T_{e},$ which is a factor (\ref{t10}) more severe.

\section{Earth gravity measurement}

In the remaining parts of the article, we describe our current results and
estimate the measurement accuracy one can achieve in ground based
experiments and in microgravity measurements. We observed MWAI using a
time-domain interferometer\cite{t12}. Two off-resonant standing wave pulses
separated by a time $T$ are applied to a sample of cold ($150\ \mu {\rm K}$) 
$^{85}$Rb atoms. The first laser pulse imposes a spatial phase modulation on
the initial atomic state, which, due to the dispersion of de Broglie waves
in free space, evolves into an amplitude modulation (representing an atomic
population grating). Owing to the Doppler dephasing,\ this grating decays in
a time of $1\mu s.$ Applying a second pulse, one restores gratings at the
echo points, 
\begin{equation}
T_{e}=(N+1)T,  \label{t12}
\end{equation}
where $N$ is an integer.

\begin{figure}[tb!]
\centering
\begin{minipage}{8.0cm}
\epsfxsize= 8 cm \epsfysize= 8.53 cm \epsfbox{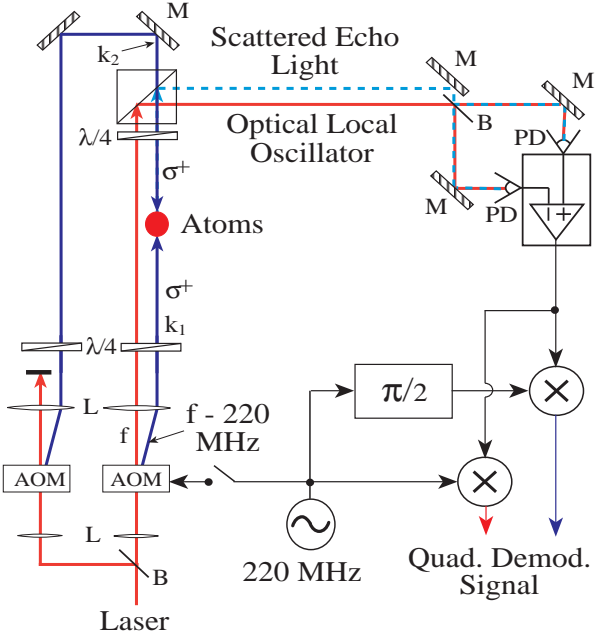}
\end{minipage}
\caption{Schematic diagram of the experimental setup. M = mirror, B = beam
splitter, PD = Photodiode, AOM = Acousto-optic modulator, $\protect\lambda %
/4=$ Quarter-wave plate, L = Lens, PBS = Polarizing Beam Splitter, and $%
\bigotimes $ = Mixer.}
\label{f04}
\end{figure}

In our experiments, $^{85}$Rb atoms are first cooled from a room-temperature
vapor in a magneto-optical trap (MOT). Approximately 12 msec after the
trapping laser beams and magnetic field are turned off (in order to allow
eddy currents to die out), the two off-resonant (between 30 and 100 MHz
detuning) standing wave pulses ($\sim 100\ $ns duration) are applied. The
standing wave pulses are composed of two traveling waves (in directions $%
{\bf {k}}_{1}$ and ${\bf {k}}_{2}$), switched on and off independently by a
pair of acousto-optic modulators (AOM). The AOMs are driven by a common
radio frequency (rf) oscillator operating at 220 MHz (see Fig.\ \ref{f04}).
The atomic grating is probed by switching on only the traveling wave along $%
{\bf {k}}_{2}$ and measuring the (complex) amplitude of the wave scattered
into the direction ${\bf {k}}_{1}$.

\begin{figure}[tb!]
\centering
\begin{minipage}{8.0cm}
\epsfxsize= 8 cm \epsfysize= 5.16 cm \epsfbox{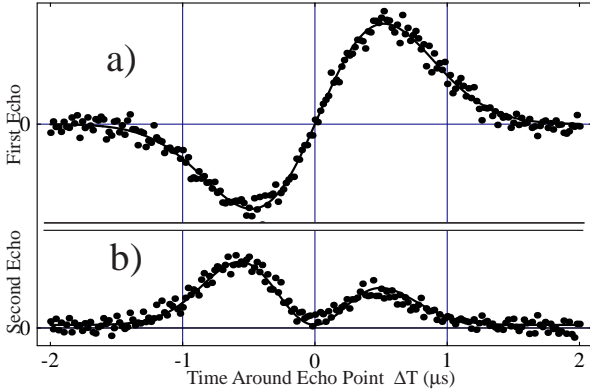}
\end{minipage}
\caption{Representative examples of the scattered signal as a function of $%
\Delta t=t$ - $T_{e}$ for (a) $T_{e}=2T$, and (b) $T_{e}=3T,$ $t$ is the total
elapsed time from the first pulse. The data in part (a) were taken using $%
\protect\sigma ^{+}$ polarized standing wave pulses, $T=799\ \protect\mu 
{\rm s}$. In (b), linearly polarized pulses were used, $T=19\ \protect\mu 
{\rm s}$.}
\label{f05}
\end{figure}
The scattered wave is detected by beating it with an optical
local-oscillator in a balanced heterodyne arrangement. The local oscillator
is derived from the light passing undiffracted through the AOM used to
switch the ${\bf {k}}_{1}$ beam. During the experiment, the echo beat signal
is further mixed down by a 220 MHz reference from the rf oscillator using a
quadrature demodulator. The two outputs of this demodulator represent the
real and imaginary parts of the scattered light field, where the real part
is in phase with the ${\bf {k}}_{1}$ field (which is \underline{not} on
during detection), and the imaginary part is $\pi /2$ out of phase with $%
{\bf {k}}_{1}$.

Typical time-dependences of the scattered signal amplitudes in the vicinity
of the echo points $T_{e}=2T$ $\left( N=1\right) $ and $T_{e}=3T$ $\left(
N=2\right) $ are shown in Fig. \ref{f05}.

\begin{figure}[tb!]
\centering
\begin{minipage}{8.0cm}
\epsfxsize= 8 cm \epsfysize= 2.48 cm \epsfbox{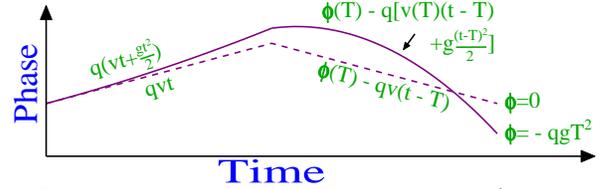}
\end{minipage}
\caption{Echo phase diagram for free space (dashed line) and in the presence
of gravity (solid line).}
\label{f06}
\end{figure}

Since our detection scheme is sensitive to the phase of the scattered
signal, it is possible to measure the influence of the gravitational
acceleration ${\bf g}$. The acceleration changes an atom trajectory and the
Doppler phase (\ref{t5pa}). The time dependence of the atom grating Doppler
phase is shown in the Fig. \ref{f06}.
We show here the process responsible for the echo at $T_{e}=2T.$ The first
pulse produces an atom grating evolving as $\exp \left[ -i{\bf q}\cdot r\left(
0\right) \right] ,$ where ${\bf r}\left( 0\right) $ is the atom position at $%
t=0.$ A nonlinear 4-quantum interaction with second pulse results in the atom
grating evolving as $\exp \left[ -i{\bf q}\cdot r\left( 0\right) +2i{\bf q}%
\cdot r\left( T\right) \right] .$ For a given space time point $\left( {\bf r%
},t\right) $ this dependence can be represented as 
\begin{equation}
\rho _{gg}\left( {\bf r}\right) \propto \exp \left( i{\bf q}\cdot {\bf r+}%
i\phi _{D}\right) ,  \label{t13}
\end{equation}
where the Doppler phase $\phi _{D}$ consists of two parts 
\begin{mathletters}
\label{t14}
\begin{eqnarray}
\phi _{D}\left( t\right) &=&\phi _{1}+\phi _{2},  \label{t14a} \\
\phi _{1} &=&{\bf q}\cdot \delta {\bf r}_{1},  \label{t14b} \\
\phi _{2} &=&-{\bf q}\cdot \delta {\bf r}_{2},  \label{t14c}
\end{eqnarray}
$\delta {\bf r}_{1}$ and $\delta {\bf r}_{2}$ are the atomic displacements
between $0$ and $T$ and between $T$ and $t.$ Without gravity $\delta {\bf r}%
_{1}={\bf v}T,\,\,\delta {\bf r}_{2}={\bf v}(t-T)$ and the phases $\phi
_{1,2}$ cancel one another at the echo point. In a gravitational field, 
\end{mathletters}
\begin{mathletters}
\label{t15}
\begin{eqnarray}
\delta {\bf r}_{1} &=&{\bf v}T+{\bf g}T^{2}/2  \label{t15a} \\
\delta {\bf r}_{2} &=&{\bf v}\left( T\right) \left( t-T\right) +{\bf g}%
\left( t-T\right) ^{2}/{2,}  \label{t15b}
\end{eqnarray}
where ${\bf v}$ is the initial velocity, ${\bf v}\left( T\right) ={\bf v}+%
{\bf g}T$. Cancellation of the Doppler phase dependence on ${\bf v}$ still
occurs at $t=2T,$ but a residual gravity dependent part $\phi _{{\bf g}}$
arises which is equal to 
\end{mathletters}
\begin{equation}
\phi _{{\bf g}}\equiv \phi _{D}\left( 2T\right) ={\bf q}\cdot {\bf g}T^{2}.
\label{t16}
\end{equation}

\begin{figure}[tb!]
\centering
\begin{minipage}{8.0cm}
\epsfxsize= 8 cm \epsfysize= 5.04 cm \epsfbox{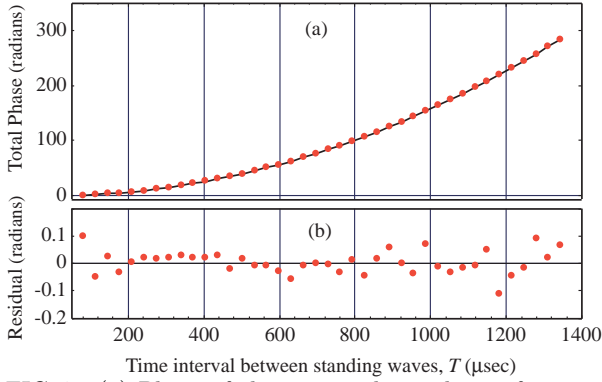}
\end{minipage}
\caption{(a) Phase of the scattered signal as a function of the pulse
spacing $T$. Points are the data and solid curve is the best fit. (b) Phase
difference between data and best fit as a function of $T$.}
\label{f07}
\end{figure}

We observe this dependence. Fig.\ \ref{f07} shows our results for the first
echo ($T_{e}=2T$) in the case where ${\bf {k}}_{1}=-{\bf {k}}_{2}$ and are
aligned to within 1 mrad of vertical, so the expected phase dependence on
the pulse spacing is $\phi _{D}\left( 2T\right) =2kg\,T^{2}$. Fig.\ \ref{f07}%
(a) shows the phase as a function of pulse spacing (points) together with a
solid curve corresponding to the best fit, which was obtained for $g=9.798$
m/s$^{2}$. Fig.\ \ref{f07}(b) shows the difference between this best fit and
the data.

\section{Recoil frequency measurement.}

The influence of atom acceleration arises whether or not the particles'
motion needs to be quantized. In contrast, effects related to recoil are
necessarily linked with quantization of the atomic center-of-mass motion 
\cite{t13}. The quantum part of dephasing (\ref{t5pb}) arises as a result of
quantization. This part leads to the periodic dependence of the echo
amplitude on the time delay between pulses. The period is given by 
\begin{equation}
\Delta T=\pi /\omega _{q}.  \label{t17}
\end{equation}
Observation of the periodical dependence allows one to measure precisely the
recoil frequency 
\begin{equation}
\omega _{q}=\hbar q^{2}/2m,  \label{t18}
\end{equation}
which is important for precise determination of the fundamental constants 
\cite{32}. The best accuracy of this measurement can be achieved in a
microgravity environment.

\begin{figure}[tb!]
\centering
\begin{minipage}{8.0cm}
\epsfxsize= 8 cm \epsfysize= 6.14 cm \epsfbox{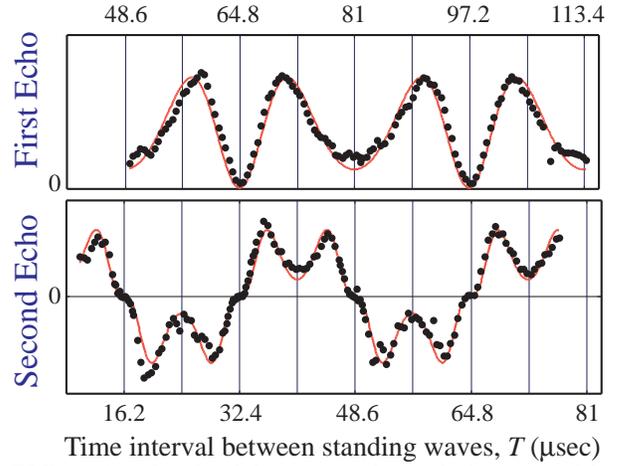}
\end{minipage}
\caption{Amplitude of the scattered signal (determined from the fits shown
in Fig.\ \ref{f05}) as a function of the pulse spacing $T$ for (a) $T_{e}=2T$
and (b) $T_{e}=3T$. The dots are the data and the solid line is the theoretical
fit\protect\cite{0,t12}.}
\label{f08}
\end{figure}

We observed \cite{t12} oscillation of the atom grating amplitudes. Examples
of the oscillating dependences are shown in Fig. \ref{f08}.
The precision $\Delta \omega _{q}$ with which one can determine the recoil
frequency is given approximately by $\Delta \omega _{q}/\omega _{q}=\Delta
T/T_{{\rm max}}$, where $\Delta T$ is the uncertainty in the time of the
zeroes of the signal and $T_{{\rm max}}$ is the maximum value of $T$ that
yields a significant signal. For our data we estimate
$\Delta \omega _{q}/\omega _{q}=(100$ ns$)/(1$ ms$)=10^{-4}$. We calculated
the period by measuring the time between well-displaced zeroes of the signal
and dividing by the number of intervening periods. This yielded a period of $%
32.388\ \mu s$ [compare this with the result of $32.386\ \mu s$ calculated
from Eq.\ (\ref{t17}) for transition $\lambda =0.78\mu $ in $^{\text{85}}$%
Rb].

\section{Ground state Ramsey fringes}

The experiments described above have been carried out in the time domain. One
can expect that the accuracy of measurement will be higher in the frequency
domain. For this purpose we propose the ground state Ramsey fringes (GSRF)
technique\cite{t10}. If traveling wave modes ${\bf k}_{1}$ and ${\bf k}_{2}$
are slightly detuned from each other by the detuning $\delta $, as
shown in Fig. \ref{f02}, the atom coherence (ground state population
grating) acquires the Ramsey phase

\begin{equation}
\phi _{R}\sim \delta T.  \label{t19}
\end{equation}
For a given time delay between pulses, one can monitor the signal from a
back-scattered field as a function of $\delta .$ The presence of atomic
acceleration or recoil affects the positions of the GSRF maxima. One obtains
another type of the acceleration or recoil effect sensor which, in contrast
to the Raman-Ramsey fringes, has the potential to increase the time of the
measurement beyond the limit imposed by the large frequency of the
involved internal transitions.

\begin{figure}[tb!]
\centering
\begin{minipage}{8.0cm}
\epsfxsize= 8 cm \epsfysize= 11.24 cm \epsfbox{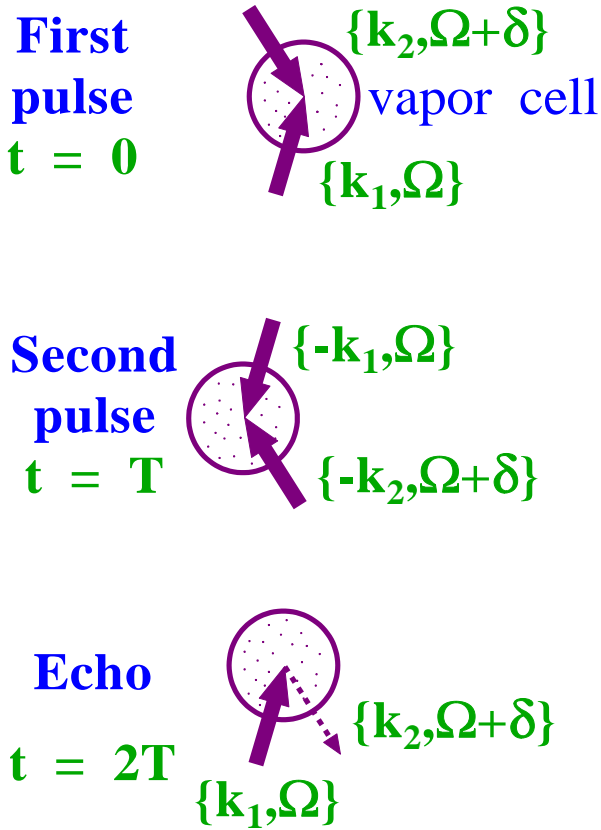}
\end{minipage}
\caption{Field geometry to observe GSRF}
\label{f09}
\end{figure}

The scheme of the GSRF is shown in fig. \ref{f09}. The first pulse produces
a grating which acquires the Ramsey phase (\ref{t19}), but it decays fast
owing to the large Doppler phase (\ref{t5pa}). For the second pulse, we
propose to reverse the directions of the travelling waves with respect to
the first pulse. If the second pulse were identical to the first, then the
Ramsey and Doppler phases would be mutually constrained: they enter into the
grating phase only in the combination $\phi _{R}-\phi _{D}$ and, therefore,
a cancellation of the Doppler phase coincides with a loss of the Ramsey
phase. But, if one reverses the directions of the travelling waves, then
non-linear processes become possible where the Doppler phases acquired
during the time intervals, $\left[ 0,T\right] $ and $\left[ T,2T\right] $,
are canceled while the Ramsey phases acquired during these intervals add to
one another.
Finally, the third, readout pulse, consisting of only one traveling mode $%
\left\{ {\bf k}_{1},\Omega \right\} ,$ scatters off the atom grating into
the field mode $\left\{ {\bf k}_{2},\Omega +\delta \right\} $ radiated by
the vapor. This signal contains the Ramsey fringes, i. e. the oscillating
dependence on the Ramsey phase (\ref{t19}).

\section{Residual gravity measurement in a microgravity environment}

The extreme sensitivity of the GSRF signal to small perturbations of the
atomic center-of-mass motion makes it ideal for acceleration measurements in
a microgravity environment. A program for carrying out such measurements
[the Space Acceleration Measurement System (SAMS)] is part of the NASA
program \cite{t14}. The sensors of residual gravity are necessary to
control the environmental quality. We propose to investigate the potential for
GSRF measurements of residual gravity in a microgravity environment.

For further estimates of the measurement accuracy, we need a minimum
measurable value of the GSRF shift $\delta .$ There is no reason for this
shift to be worse than that achieved by the Raman-Ramsey technique, but, as
we have explained above, for GSRF this shift can be smaller. The ultimate
limit for GSRF shift will be established by experiments, either ground based
or gravity free. In this article we take the lower boundary of the
shift equal to 
\begin{equation}
\delta _{\min }=2\pi \,15\,mHz.  \label{t21}
\end{equation}
One can get this value as a product of the relative accuracy $\left(
1.1\,10^{-7}\right) $ of the recoil frequency measurement using the
Raman-Ramsey technique \cite{31}\ and $\omega _{q}$ $\left( 2\pi
\,130\,kHz\right) .$ The halfwidth of the Raman-Ramsey fringes in the
experiment of Ref.\cite{31} was 
\begin{equation}
\Gamma _{R}\approx 2\pi \,8\,Hz.  \label{t22}
\end{equation}

Let us estimate the minimum residual acceleration ${\bf g}_{r\min }$ which
can be measured by GSRF in a cold $^{85}$Rb vapor or beam. If GSRF in time
separated fields are used, the gravitational phase shift $\phi _{g_{r}}$ in
the vicinity of the $t=2T$ echo point is obtained by replacing ${\bf %
g\rightarrow g}_{r}$ in Eq. (\ref{t16}), $\phi _{g_{r}}={\bf q\cdot g}%
_{r}T^{2}.$ When the phase shift is small, it results in a shift of the GSRF
by an amount $\delta _{g_{r}}\sim \phi _{g_{r}}/T={\bf q\cdot g}_{r}T.$
Since we assume that a frequency measurement with accuracy (\ref{t21}) is
possible, for ${\bf q\parallel g}_{r}$ and $T=100\,ms$ one gets for a lower
limit of the residual gravity measurement 
\begin{equation}
g_{r\min }\approx 6\,10^{-9}g.  \label{45}
\end{equation}

To measure such a small residual gravity, one has to exclude corrections
arising from rotations. Rotation with frequency $\Omega _{r}$ leads to a
Coriolis acceleration of order of $2\left\langle u\right\rangle \Omega _{r},$
where $\left\langle u\right\rangle $ is the mean atomic velocity. In an
atomic beam launched with velocity $u=17\,cm/s,$ the requirement that the
Coriolis acceleration produce a shift less than the minimum residual
gravitational shift given by Eq. (\ref{45}) can be stated as 
\begin{equation}
\Omega _{r}<3\,10^{-3}\Omega _{earth}.  \label{47}
\end{equation}

The situation improves if one uses a cold atomic vapor, such as that used in
our experiment\cite{t12} and in experiments on optical transitions in Mg\cite
{15}. The corrections from rotation would vanish identically if the mean
velocity of the vapor were identically equal to zero. If there is some
asymmetry in the distribution that gives rise to a mean velocity $%
\left\langle u\right\rangle =\alpha \Delta u,$ where $\Delta u$ is the width
of the velocity distribution, then it is necessary that 
\begin{equation}
\alpha <g_{r\min }/2\Delta u\Omega _{r}.  \label{48}
\end{equation}
for rotational effects to be negligible. Even if the space station rotates
with the earth's rotation rate, for the transition in $^{85}$Rb and $\Delta
u=17\,cm/s,$ one finds that $\alpha <3\,10^{-3}$ is needed$.$ This less
severe restriction shows that it is better to use cold vapors than beams for
precise measurements of residual gravity.

\section{Atom gyroscope}

In this section we estimate the precision of an atomic gyroscope based on
GSRF. Estimates have to be performed separately for the microgravity
environment and ground-based experiment.

\subsection{Atom gyroscope in a microgravity environment.}

We propose to use GSRF as a sensor of a system's rotation. When Ramsey
fringes are observed in a frame rotating with angular frequency ${\bf \Omega 
}_{r}$, the atoms undergo a Coriolis acceleration equal to 
\begin{equation}
{\bf a=}2{\bf v\times \Omega }_{r}.  \label{11}
\end{equation}
Owing to this acceleration, a new type of dephasing $\phi _{r}$ arises\cite
{5}. To obtain this dephasing for a small rotation rate, one can neglect
changes of the acceleration resulting from changes in the atomic velocity.
In this limit, one can simply replace ${\bf g}$ by ${\bf a}$ in Eq. (\ref
{t16}) to obtain the rotational dephasing $\phi _{r}$ given by 
\begin{equation}
\phi _{r}=2\left( {\bf q\times \Omega }_{r}\right) \cdot {\bf v}T^{2}.
\label{12}
\end{equation}
A direct observation of this fringe shift has been reported previously using
microfabricated structures to scatter atoms \cite{19,22}. A precision of 
\begin{equation}
\delta \Omega _{r}=0.042\text{ }\Omega _{earth}=0.63\,%
{{}^\circ}%
/h,  \label{13}
\end{equation}
where $\Omega _{earth}=15%
{{}^\circ}%
/h$ is the earth rotation rate, has been achieved \cite{22}. We have already
discussed rotational observations using optical Ramsey fringes 
\cite{6} and Raman-Ramsey fringes\cite{Gus97,t9}. In the latter case the
precision is 
\begin{equation}
\delta \Omega _{r}=6\text{\thinspace }10^{-3}\Omega _{earth}=0.08\,%
{{}^\circ}%
/h.  \label{t20}
\end{equation}
This precision is still below that of fiber-optic-gyroscopes \cite{23} $%
\left( 0.047\,%
{{}^\circ}%
/h\right) $, but already better than that of ring-laser gyroscopes\cite{23p} $%
\left( 0.18\,%
{{}^\circ}%
/h\right) $.

\begin{figure}[tb!]
\centering
\begin{minipage}{8.0cm}
\epsfxsize= 8 cm \epsfysize= 2.18 cm \epsfbox{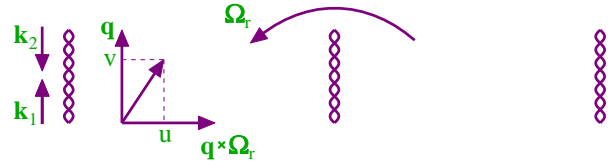}
\end{minipage}
\caption{Gyroscope based on ground state Ramsey fringes.}
\label{f10}
\end{figure}

One can use the Ramsey resonance shift as a rotation sensor\cite{5,6}. In
contrast to measurements of photon recoil effects or gravitational
acceleration which can be measured using either temporally or spatially
separated pulses, gyroscopic measurements involving frequency measurements
should be carried out using spatially separated fields\cite{5}. A scheme for
accomplishing this using GSRF is shown in Fig.\ \ref{f10}.
If the rotation frequency is perpendicular to ${\bf q}$, the rotation phase (%
\ref{12}) is equal to $\phi _{r}=2q\Omega _{r}sign\left( u\right) L^{2}/u$,
where $L=uT$ is the spatial separation of the fields. The Ramsey phase $\phi
_{R}=2\delta T=2\delta L/u$ has the same dependence on longitudinal velocity 
$u$ as $\phi _{r}$. The total phase is 
\begin{equation}
\phi =\phi _{R}+\phi _{r}=2\left( \delta +qL\Omega _{r}\right) /u.
\label{14}
\end{equation}
The Ramsey fringes are centered at the point where $\phi =0$. As a result of
the rotation, the central fringe is shifted by 
\begin{equation}
\delta =-sign\left( u\right) qL\Omega _{r}.  \label{15}
\end{equation}

The highest gyroscope accuracy can be reached in a microgravity
environment. If an effusive beam of atoms with thermal velocity $u$ crosses
three interaction zones, each separated from one another by distance $L,$
than the Ramsey fringes halfwidth is given by \cite{1pp} 
\begin{equation}
\Gamma _{R}\approx 0.6u/L.  \label{t23}
\end{equation}
The distance between fields $L$ can be chosen arbitrarily. A large distance
allows for a more precise measurement of the rotation frequency. We set $%
L=20\,$cm in arriving at our estimates$.$ For the fringe halfwidth (\ref{t22}%
) and $u\approx 20\,m/s$ on a
transition in $^{\text{85}}$Rb $\left( \lambda =0.78\mu
,q=1.6\,10^{5}\,cm^{-1}\right) $ with the shift given by Eq. (\ref{t21}), the
minimum measurable rotation rate is equal to 
\begin{equation}
\Omega _{r\min }\sim 4\,10^{-4}\Omega _{earth}\approx 6\,10^{-3}%
{{\,\,}^\circ}%
/h.  \label{t24}
\end{equation}

\subsection{GSRF gyroscope in ground based measurements}

In principle, the same accuracy (\ref{t24}) can be achieved in ground
based experiments, but for this purpose the wave vectors ${\bf k}_{i}$ have
to be aligned in the horizontal plane to exclude gravitational
influence. Let us estimate the requirements for this alignment. The typical
time of flight of atoms between separated fields is $T\sim L/u\sim 1.5$/$%
\Gamma _{R}\approx 30\,ms.$ If $\theta $ is the order of magnitude of
the small residual angle between the wave vectors and the horizontal plane,
then the gravitational phase (\ref{t16}), $\phi _{g}\sim qg\theta T^{2},$
leads to the GSRF shift $qg\theta T$. When we require this to be smaller
than the minimum measurable shift (\ref{t21}), one finds that for a
transition in $^{85}$Rb, 
\begin{equation}
\theta \lesssim \frac{\lambda \delta _{\min }}{4\pi gT}\approx
2\,10^{-8}\,rad,  \label{t25}
\end{equation}
which is two orders more severe than the present state-of-the-art limit for
alignment \cite{46}$.$

\begin{figure}[tb!]
\centering
\begin{minipage}{8.0cm}
\epsfxsize= 8 cm \epsfysize= 2.31 cm \epsfbox{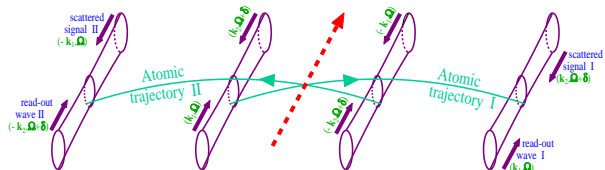}
\end{minipage}
\caption{A scheme of the atomic gyroscope in the presence of the earth's
gravity field.}
\label{f11}
\end{figure}

For the further elimination of gravity in ground based measurements, one can observe
simultaneously two scattered signals, I and II (see Fig. 
\ref{f11}). These signals both contain GSRF's, whose shifts $\delta _{I,II}$
consist of gravitational and rotational parts $\delta _{I,II}^{g}$ and $%
\delta _{I,II}^{r}$. From Fig.\ \ref{f11}, one concludes that signal II is
produced by the fields consisting of traveling wave field modes $\left( \mp 
{\bf k}_{2},\Omega +\delta \right) $ and $\left( \mp {\bf k}_{1}{\bf ,}%
\Omega \right) $ instead of modes $\left( \pm {\bf k}_{1}{\bf ,}\Omega
\right) $ and $\left( \pm {\bf k}_{2},\Omega +\delta \right) $ responsible
for signal I (signal II is obtained from signal I by the substitutions ${\bf %
k}_{1}{\bf \leftrightarrow -k}_{2},$ $\Omega \leftrightarrow \Omega +\delta
) $. Under this transformation the wave vector difference ${\bf q}$ given by
Eq. (\ref{20}) is not changed, but the detuning $\delta $ changes sign.
Thus, the gravity-induced parts of the shifts have opposite sign, $%
sign\left( \delta _{II}^{g}\right) =-sign\left( \delta _{I}^{g}\right) ,$
and the sum of the shifts $\delta =\delta _{I}+\delta _{II}$ is gravity
insensitive {\em provided }that the fields are imposed symmetrically with
respect to the symmetry axis of the atomic spatial-velocity distribution
(dashed arrow on the Fig. \ref{f11}). A similar idea for eliminating the
effects of gravity in rotation measurements has been considered recently\cite
{47} and realized in the Raman-Ramsey gyroscope\cite{t9}.

This severe requirement for the atomic phase space distribution to have a
symmetry axis is easier to realize for atoms in a cell in thermal equilibrium than in
counterpropagating atomic beams.

\section{Higher order atomic gratings.}

\begin{figure}[tb!]
\centering
\begin{minipage}{8.0cm}
\epsfxsize= 8 cm \epsfysize= 5.51 cm \epsfbox{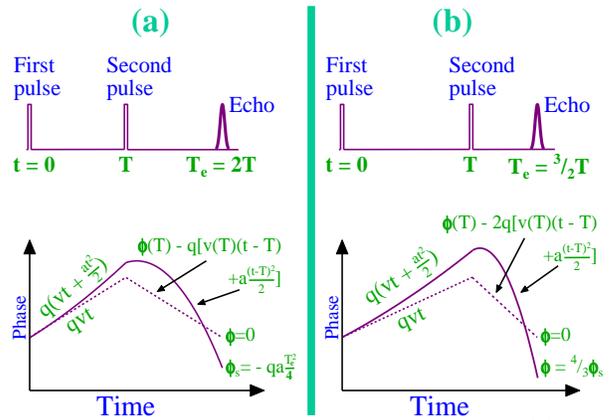}
\end{minipage}
\caption{Echo phase diagrams corresponding (a) to the first-order grating
localized near the echo point $T_{e}=2T$ \ and (b) second order grating
localized near $T_{e}=\,^{3}/_{2}T$ for free space (dashed line) and in the
presence of gravity (solid line).}
\label{f12}
\end{figure}

To increase the sensitivity to an atom's acceleration, one can use higher
order atom gratings having wave numbers that are a multiple of $q.$ The nonlinear
interaction of an atom with a pulse of non-copropagating fields applied to
the same transition leads to higher order grating production in the atom
ground state population\cite{0,1ppp,1pp,3,4,8,t10,t11,t12}. This is in
contrast to the interaction with separated traveling waves\cite{5,6,15} or
counterpropagating traveling waves applied on adjacent transitions\cite
{30,31,32,20,Gus97,t9,47}, where to get higher harmonics one needs a number
of pulses\cite{31}. (The only exception is a multiple beam atom
interferometer\cite{t15}.) When higher harmonics are involved in the process
of grating formation, one can expect that the part of the Doppler phase (%
\ref{t16}) caused by acceleration ${\bf a}$ increases. This should allow one
to increase the accuracy of the acceleration measurement. This statement is
illustrated in Fig. \ref{f12}, where the phases of the first and second
order grating [evolving as $\cos \left( {\bf q}\cdot {\bf r}\right) $ and $%
\cos \left( 2{\bf q}\cdot {\bf r}\right) $, respectively] are compared.
Consider the general case. If the first strong pulse produces an atom
grating evolving as $e^{in_{1}{\bf q}\cdot {\bf r}}$, this grating acquires
the Doppler phase part, 
\begin{equation}
\phi _{1}=n_{1}{\bf q}\cdot \delta {\bf r}_{1},  \label{t26}
\end{equation}
where $\delta {\bf r}_{1}$ is the atomic displacement, during the time
interval $\left[ 0,T\right] .$ Similarly, if after the second pulse the
grating $e^{in{\bf q}\cdot {\bf r}}$ is produced, it acquires an additional
phase during the time interval $\left[ T,t\right] $, 
\begin{equation}
\phi _{2}=n{\bf q}\cdot \delta {\bf r}_{2},  \label{t27}
\end{equation}
where $\delta {\bf r}_{2}$ is the corresponding atomic displacement. Using
expressions (\ref{t15}) one finds that the total phase $\phi _{D}(t)$ is
given by 
\begin{eqnarray}
\phi _{D}(t) &=&{\bf q}\cdot {\bf v}\left[ n_{1}T+n\left( t-T\right) \right]
+  \nonumber \\
&&^{1}/_{2}{\bf q}\cdot {\bf a}\left[ n_{1}T^{2}+n\left( t^{2}-T^{2}\right) %
\right] .  \label{t28}
\end{eqnarray}
The echo point is by definition, a time, $t=T_{e},$ where the velocity
dependent part is canceled. One find that this cancellation occurs at all
points 
\begin{equation}
T_{e}=\frac{n^{\prime }}{n}T,  \label{t29}
\end{equation}
where $n^{\prime }=n-n_{1}$ $\left( \left| n^{\prime }\right| >\left|
n\right| \right) ,$ integers $n^{\prime }$ and $n$ have no common factor.
Since after the second pulse the grating evolves as $e^{in{\bf q}\cdot {\bf r%
}},$ the grating period is given by $2\pi /nq.$ One sees that the first
harmonic $\left( n=1\right) $ is localized at the points $2T,$ $3T,\ldots $,
the second harmonic $\left( n=2\right) $ is localized at $^{3}/_{2}T,$ $%
^{5}/_{2}T,\ldots $ and so on.

At the echo point the residual Doppler phase $\phi _{{\bf a}}\equiv \phi
_{D}(T_{e})$ associated with acceleration ${\bf a}$ can be represented as 
\begin{equation}
\phi _{{\bf a}}=2n(1-T/T_{e})\phi _{s},  \label{t30}
\end{equation}
where $\phi _{s}={\bf q}\cdot {\bf a}T_{e}^{2}/4$ the minimum value of $\phi
_{{\bf a}}$ at the point $T_{e}=2T.$ The echo time $T_{e}$ is evidently the
total time of the atomic coherence evolution. To compare
acceleration-related phases, this time has to be fixed. The ratio $\phi _{%
{\bf a}}/\phi _{s}$ for the different grating periods and different time
separation between pulses $T$ is shown in the following table 
\[
\begin{array}{ccc}
\begin{array}{c}
Grating \\ 
period
\end{array}
& 
\begin{array}{c}
Time\,\,of \\ 
evolution
\end{array}
& \frac{\phi _{a}}{\phi _{s}} \\ 
2\pi /q & T_{e}/T=2 & 1 \\ 
2\pi /q & T_{e}/T=3 & 4/3 \\ 
2\pi /q & T_{e}/T=\infty  & 2 \\ 
\pi /q & T_{e}/T=3/2 & 4/3 \\ 
\pi /q & T_{e}/T=5/2 & 12/5 \\ 
\pi /q & T_{e}/T=\infty  & 4 \\ 
2\pi /3q & T_{e}/T=4/3 & 3/2 \\ 
2\pi /3q & T_{e}/T=7/3 & 24/7 \\ 
2\pi /3q & T_{e}/T=\infty  & 6
\end{array}
\]

\section{Three-pulse echo technique.}

In this section we describe our recent technique \cite{t16} for observing
the higher order atomic gratings. Higher order gratings allow one to improve
the sensitivity of inertial measurements.

In the Sec. III we described the back-scattering technique to observe atomic
grating.\ The grating was detected by applying a traveling wave to the
atomic cloud and measuring the coherent backscattering of this traveling
wave. One of the properties of this detection technique is that it is
sensitive only to the second harmonic of the atomic density distribution
(i.e. the spatial Fourier component with period $2\pi /q$). This detection
technique has the nice feature for atom interferometry experiments that the
signal is not accompanied by a (possibly large) background due to the zeroth
harmonic of the atomic density distribution. It has a disadvantage in that
it reveals no direct information about the higher harmonics
(4th, 6th, etc).

By applying a second standing wave at a time $T_{2}$ after the initial
standing wave, one can rephase the 1st order grating (with periodicity $2\pi
/q$) and produce high ($n$th) order gratings (with periodicity $2\pi /nq$)
at various times after the this second pulse. To observe higher order
gratings, we apply a third standing-wave pulse (SW3) whose purpose is to
convert the higher-order grating into a 2nd-order grating that can be
detected by the back-scattering technique described above.

\begin{figure}[tb!]
\centering
\begin{minipage}{8.0cm}
\epsfxsize= 8 cm \epsfysize= 4.87 cm \epsfbox{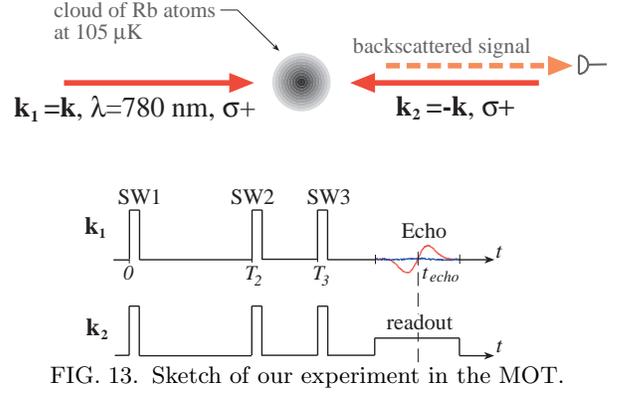}
\end{minipage}
\caption{Sketch of our experiment in the MOT.}
\label{f13}
\end{figure}

\begin{figure}[tb!]
\centering
\begin{minipage}{8.0cm}
\epsfxsize= 8 cm \epsfysize= 8 cm \epsfbox{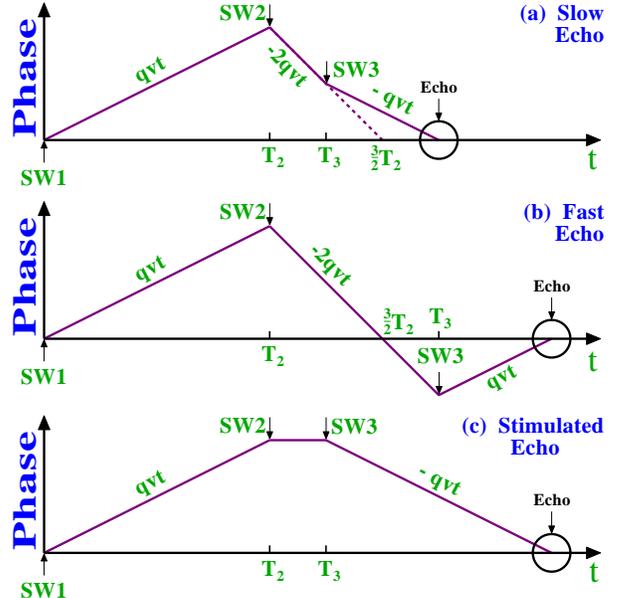}
\end{minipage}
\caption{Phase diagrams of three different three-pulse echoes.}
\label{f14}
\end{figure}

\begin{figure}[tb!]
\centering
\begin{minipage}{8.0cm}
\epsfxsize= 8 cm \epsfysize= 6.02 cm \epsfbox{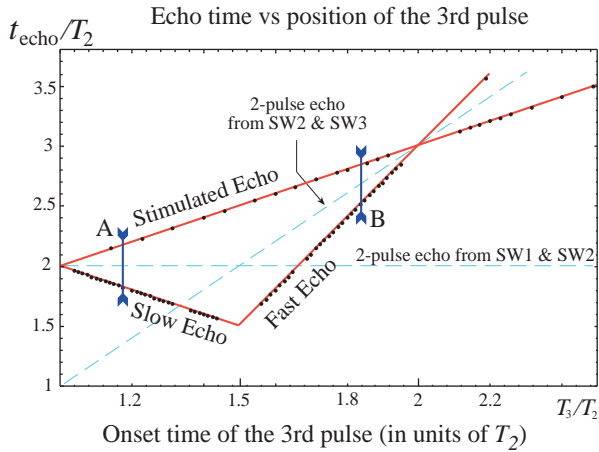}
\end{minipage}
\caption{In this echo diagram, solid lines represent the echos that have
been observed. The data points pertain to the three-pulse echo that we wish
to study.}
\label{f15}
\end{figure}

Our experiments were carried out with a cloud of $^{85}$Rb atoms cooled down
to 105 $\mu $K in a magneto-optical trap. The cloud was illuminated by a
series of three optical pulses of two $\sigma ^{+}$-polarized plane waves
traveling in opposite directions, ${\bf k}_{1}$ and ${\bf k}_{2}$ (see
Figure \ref{f13}).
The frequency of the optical pulses was detuned above the atomic transition
frequency of the $^{85}$Rb 5S$_{1/2}\,(F=3$) -- 5P$_{3/2}\,(F^{\prime }=4$)
transition by 400 MHz. The role of spontaneous processes was therefore
negligible. The first pulse (SW1) creates a spatial grating in the atomic
cloud. The grating rapidly vanishes due to the Doppler dephasing. The pulse
SW2 causes the reappearance of the gratings of various orders at later times
(see Fig.~\ref{f14}).
At a time in the vicinity of $(3/2)T_{2}$, where the 2nd order grating is
expected to occur, we applied a third standing wave pulse. Figures \ref{f14}
(a) and (b) show the effect of the third pulse on the second order grating
(period $\pi /q$). Echoes occur at times when the lines cross the horizontal
($t$) axis. If the third pulse occurs {\em \ before} $(3/2)T_{2}$, we call
the signal the {\it slow} echo and observe the echo shown in Fig.~\ref{f14}%
(a). On the other hand, Fig.~\ref{f14}(b) shows the situation when the third
pulse occurs {\em after} the time $(3/2)T_{2}$, which we call the {\it fast}
echo. In addition, we find another echo signal that depends on all three of
the standing-wave pulses, shown in Fig.~\ref{f14}(c). These echoes can be
distinguished experimentally by measuring their time of occurrence as a
function of the time of application of the excitation pulses. In particular,
Figure \ref{f15} shows the time of the echoes of Fig.~\ref{f14} as a
function of the time of the third pulse.
\begin{figure}[tb!]
\centering
\begin{minipage}{8.0cm}
\epsfxsize= 8 cm \epsfysize= 7.19 cm \epsfbox{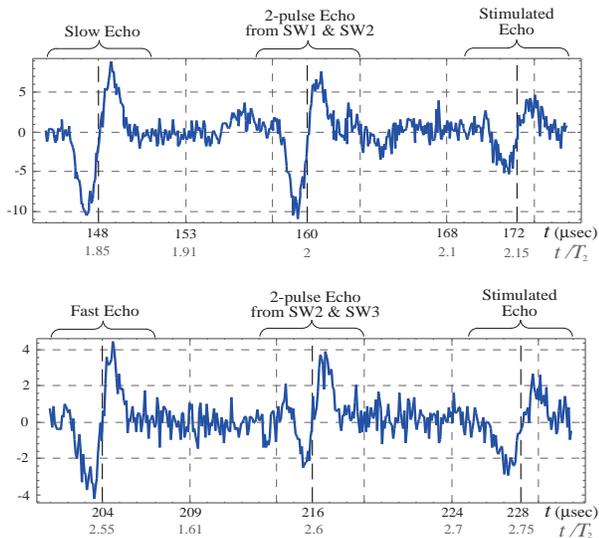}
\end{minipage}
\caption{Echo signals. The two time traces correspond to A and B in Figure~%
\ref{f15}.}
\label{f16}
\end{figure}
Around the echo time, $T_{e}$, the grating is probed by a long weak pulse
with traveling wave vector ${\bf k}_{2}$. The grating backscatters in the
direction ${\bf k}_{1}$. The scattered light strikes a heterodyne detector
sensitive to the field amplitude and phase. The dots in figure \ref{f15} are
the actual experimental data points representing times when echo signals
were detected. The vertical lines marked A and B correspond to two time
traces shown in figure \ref{f16}.
In these traces one can clearly see three different echoes, the one
two-pulse and two three-pulse echoes, that fall into the selected time
interval. It is interesting to notice that when the SW3 pulse is applied,
the amplitude of the two-pulse echo produced by SW1 and SW2 becomes smaller.
This is because a part of the phase trajectories leading to this echo is
converted by SW3.

\begin{figure}[tb!]
\centering
\begin{minipage}{8.0cm}
\epsfxsize= 8 cm \epsfysize= 6.2 cm \epsfbox{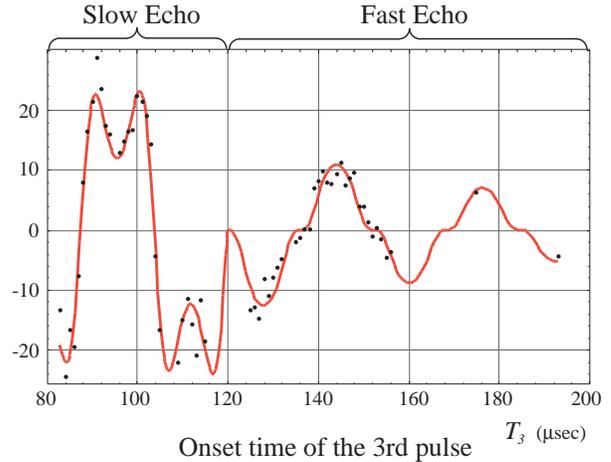}
\end{minipage}
\caption{Dependence of the fast and slow echos amplitude on $T_{3}$.}
\label{f17}
\end{figure}

The contrast of the higher order grating can be found from the echo
amplitude. We have studied the amplitude of the fast and slow echoes as a
function of $T_{3}$. The experimental data and theoretical fit are shown in
figure \ref{f17}.
Even though the theory fits the experiment well, additional study is
required to be able to monitor the higher order gratings by looking at the
echo signal.

To summarize, in our three-pulse experiment we successfully detected the
second order atomic density grating in a MOT cloud. In addition, use of
three standing wave pulses allows us not only to detect the higher order
gratings, but also to convert them to yet higher orders. It is also a first
step in our study of multi-pulse echoes. One intention is to measure
gravitational acceleration using higher order gratings.

\section{Conclusion}

Matter wave interference techniques can lead to significant improvements in
the precise measurement of inertial, gravitational, and quantum effects.

Being insensitive to the energy of the atomic internal motion, the shape and
the center of the ground state Ramsey fringes do not depend on perturbations
of the internal state. Simultaneously, the accuracy of measurements is not
related to any frequency of internal transitions. It allows one to increase
the time of the atomic coherences' evolution $T_{e}$ by 2-3 orders of
magnitude$,$ or reduce requirements for frequency stabilization by 2-3
orders of magnitude.

These advantages can be exploited most fully in a microgravity environment,
where the experiments do not suffer the enormous gravitational dephasing of
atom coherences between different center-of-mass states.

\acknowledgments

It is a pleasure to acknowledge useful discussions with J. L. Cohen and A.
V. Turlapov, and their help in the manuscript preparation. This research is
supported by the U. S. Army Research office under grant number
DAAG5-97-0113, by the National Science Foundation under grants PHY-9414020
and PHY-9800981, NYU and the Packard Foundation..

\end{multicols}

\begin{references}
\bibitem{1p}  Owing to relativistic corrections, dephasing caused by the
second order Doppler effect also arises. This has been observed and studied
independently in Refs. \cite{1ppp,1pp}.
\end{references}
\end{document}